# Observation of Multiphoton-induced Fluorescence from Nano Graphene Oxide and Its Applications in *In vitro* and *In vivo* Bioimaging**

Jun Qian[1,+], Dan Wang[1,+], Li Peng[2], Wang Xi[3], Fu-Hong Cai[1], Zhen-Feng Zhu[1], Hao He[4], Ming-Lie Hu[4], and Sailing He[1,]*

[1] Centre for Optical and Electromagnetic Research, State Key Laboratory of Modern Optical Instrumentation (Zhejiang University), Zhejiang University, Joint Research Center of Photonics of the Royal Institute of Technology (Sweden), Lund University (Sweden) and Zhejiang University (ZJU), Hangzhou, Zhejiang, 310058 (China).

[2] ZJU-SCNU Joint Research Center of Photonics, South China Normal University (SCNU), Guangzhou, 510006, China

[3] Department of Neurobiology, Key Laboratory of Medical Neurobiology of Ministry of Health of China, Zhejiang Province Key Laboratory of Neurobiology, School of Medicine, Zhejiang University, Hangzhou, Zhejiang, China.

[4] Ultrafast Laser Laboratory, College of Precision Instrument and Optoelectronics Engineering, Key Laboratory of Opto-Electronics Information and Technical Science, Ministry of Education, Tianjin University, Tianjin 300072, China

[+]Jun Qian and Dan Wang contributed equally to this work.

*Corresponding author: sailing@kth.se

**Note:** We submitted this manuscript to *Angew. Chem. Int. Ed.* on Jan 25, 2012. We started this independent work about ten months ago. Just before the submission of the present manuscript, we noticed a related work, which is on two-photon (but not three-photon) bio-imaging (but not for in vivo or subcellular components), has just been published in the same journal (DOI: 10.1002/anie.201106102).

Carbon nanomaterials typically include zero-dimensional (0D) fullerene, 1D carbon nanotubes (CNTs) and 2D graphene. Since it was first discovered and observed by Gein's group in 2004,[1] graphene has consistently been a hot research topic due to its unique electronic and mechanical properties.[2] Graphene also possesses special optical performance (e.g., saturable absorption, surface enhanced Raman scattering), and has already exhibited its wide application potentials in e.g. mode locking pulsed laser, solar cells and molecule sensing.[3] However, graphene has rarely been applied in bio-photonics research, which is mainly due to its drawback of non-dispersity in water. As a derivative of graphene, graphene oxide (GO) has many advantages in various areas of bioimaging and bio-therapy research.[4] Firstly, GO is mainly comprised of carbon elements, and it is non-toxic and more biocompatible[5] than other existing nanoparticles such as heavy metal ion (e.g., quantum dot), noble metal (e.g., gold/silver nanoparticle) and rare earth ion (e.g., upconversion nanocrystal).[6] GO is usually synthesized by partially oxidizing graphene, and the formation of graphitic islands on its surface is expected to produce quantum confinement effect, which can open up a band gap in graphene, as well as make GO possess photonluminescence performance.[7] Single-layered "thin" GO has two sides exposed on the surface, which can make it achieve a higher drug/bio-molecule loading efficiency compared with CNTs and other nanoparticles. Furthermore, GO also has a distinct photothermal effect, which is similar to some noble metal nanoparticles. Due to these unique features, the use of GO has been increasingly investigated in recent years in *in vitro* bio-imaging as a luminescent probe, anticancer drug delivery as a molecule-carrier, photodynamic therapy of tumor as a photon-thermal transducer.[8]

Although research on single photon (or downconversion) induced photonluminescence and two-photon absorption-only behaviors of GO have already been presented,[9] to the best of our knowledge its multiphoton (or upconversion) induced photonluminescence performance has never been reported. Multiphoton excitation induced bio-imaging has many unique advantages.[10] Due to a square/cubic or higher dependence of multiphoton absorption on laser intensity, the sample region outside the beam focus cannot be excited, and it could reduce the possibility of photobleaching of the sample signal.[11] The nonlinear excitation mode is also helpful to improve the spatial resolution of imaging, since only the site where the laser beam is focused can be efficiently excited. More importantly, multiphoton excitation has great potentials for deep-range tissue imaging. For one-photon bioimaging, photosensitizers usually absorb light in the visible spectral region below 700 nm, where light penetration into the tissue is limited. On the other hand, the excitation laser wavelength is usually in the range of 700-900 nm for two-photon excitation, and in the range of 1000-1350 nm for three-photon excitation. The former is typically considered as a main transparent window of light for tissues. The latter opens a second window for *in vivo* imaging with low tissue autofluorescence, and consequently the penetration of multiphoton excitation light and the generation of deep-tissue signals can be improved.[12] Femtosecond (fs) pulsed laser is a promising light source for multiphoton excitation. With low average power, high peak power and relatively low repetition period, a fs laser can be properly operated to produce a mimimal thermal effect during its interaction with bio-samples, which can effectively avoid certain thermal damage. Under imaging guiding, a fs laser can also achieve cellular/subcellular behavior manipulation, e.g., cell fusion, gene transfection, and $Ca^{2+}$ influx control.[13]

In the present paper, we observed both two-photon and three-photon induced distinct photoluminescence from GO nanoparticles under fs laser excitation. Conjugated with PEG molecules, GO nanoparticles exhibited high chemical stability, and could effectively label HeLa cells. Imaged with a two-photon scanning microscope, GO nanoparticles were observed to localize in the mitochondria, endoplasmic reticulum, Golgi and lysosome of HeLa cells. Furthermore, GO nanoparticles were micro-injected into the brain of a black mouse, and *in vivo* two-photon luminescence imaging illustrated that GO nanoparticles located at 300 μm depth in the brain could be clearly distinguished.

GO nanoparticles were synthesized from bulk graphite with a modified Hummer's method (Figure 1a). According to TEM analysis (Figure 1b), the as-synthesized GO nanoparticles have an average size of about 40 nm. An atom force microscope (AFM) picture illustrated that the average thickness of the as-synthesized GO nanoparticles was about1 nm (Figure 1c),[14] which was close to that of single-layer GO nanoparticles. Aqueous solution of GO nanoparticles was mixed with gold nanorods (GNRs) and then added into a quartz cuvette. The enhanced Raman spectrum of GO nanoparticles was then measured under a 785 nm laser excitation. Graphite with pure $sp^2$ carbon atoms usually has a characteristic Raman peak around 1585$cm^{-1}$, which is called G band, and its intensity is closely related to the size of graphite crystal. However, the as-synthesized GO nanoparticles have two distinct Raman peaks. One still located around 1585$cm^{-1}$ while the other appeared at 1331$cm^{-1}$ (Figure S1). The latter is called D band and is related with $sp^3$ carbon atoms. The Raman spectrum indicated that part of $sp^2$ carbon atoms have been transferred to $sp^3$ structure during graphite oxidation process, which is a typical characterization of GO structure.[15] Extinction spectrum of aqueous solution of GO nanoparticles was measured by a UV-vis scanning spectrophotometer. Only a scattering spectrum without a distinct absorption band could be observed (Figure 1d), and it mainly arose from the Rayleigh scattering by GO nanoparticles (with an average diameter of 40nm). Since the scattering intensity is inverse proportional to $\lambda^4$ ($\lambda$ is the optical wavelength), the scattering-induced extinction of GO nanoparticles in a short wavelength (e.g., 300-450 nm) was more distinct than that in a long wavelength. One-photon luminescence performance of GO nanoparticles was then measured by a Fluorescence Spectrophotometer with an excitation wavelength of 405 nm. The photoluminescence spectrum is in the range of 450 to 700 nm, with the maximum located at 550 nm (Figure 1d). The excitation spectrum of GO nanoparticles was measured with emission wavelength set at 550 nm and shown in Figure S2. We also noticed that the one-photon luminescence spectra of GO nanoparticles under various excitation wavelengths were almost the same (Figure S3).

According to one-photon excitation and luminescence spectra of GO nanoparticles, a fs laser with a wavelength in the 700-900 nm range can be used for two-photon excitation. Figure S4 shows the linear transmission spectra of 1-cm-thick layer of water and aqueous dispersion of GO nanoparticles, and both of them had negligible one-photon attenuation at 700-900 nm range. We used a 810 nm, ~160 fs pulsed laser at a repetition rate of 1 kHz for two-photon study of aqueous solution of GO nanoparticles in cuvette, and the laser beam was from an optical parametric generator (OPG) pumped by a Ti-sapphire oscillator/amplifier system (Coherent, Inc.). The luminescence spectrum was recorded with an optical fiber spectrometer (PG2000, Ideaoptics Instruments). As shown in Figure 2a, the two-photon luminescence spectrum of GO nanoparticles is in the range of 450 to 700 nm (with the maximum located at 550 nm), and its envelope is very similar to that under one-photon excitation. That means in both one-photon and two-photon processes, the excited GO nanoparticles were finally relaxed to the same lowest excited electronic-vibrational state(s), from which the luminescence emission occurred. Thus, when the GO sample was stimulated by 810 nm laser pulses, it needs to absorb at least two photons at the same time to get excited (Figure S5a).[16] To verify that the observed luminescence was induced by two-photon excitation, we measured series of emission spectra of GO nanoparticles by fs laser (at 810 nm) with various average powers (Figure 2a). The luminescence intensity dependence on the pump energy is plotted and shown in Figure 2b, where one can see that the peak emission intensity is proportional to the square of the pump pulse energy, confirming the characterization of a two-photon process. We then attempted to measure the two-photon excitation spectrum of GO nanoparticles from 700 to 900 nm. To achieve the same two-photon luminescence intensity, fs laser of different wavelengths with different average powers were utilized to pump the sample. We selected 21 wavelengths in the range of 700 to 900 nm (10 nm per point), and a fitted function of two-photon excitation spectrum was obtained in Figure 2c. The measured two-photon excitation efficiency is higher in the range of 740-820 nm (as the product of the nonlinear two-photon absorption cross section and the luminescence quantum efficiency is higher), illustrating that less laser power in this wavelength range is needed to achieve sufficient luminescence signals.[17] Aqueous solution of GO nanoparticles with original and diluted concentrations were dropped on the glass slides and imaged under an upright two-photon scanning microscope (Olympus, FV1000, fs laser at 810 nm) separately. A 3D reconstructed picture illustrating the intensity dependence on the X-Y position of the original GO sample is shown in Figure S6. A luminescence image of a diluted GO sample can be observed in the inset of Figure 2d, while its spectrum (Figure 2d) coincided well with the one measured in a cuvette.

We then used a 1215 nm, ~160 fs pulsed laser at a repetition rate of 1 kHz from the same OPG system to perform the measurement of three-photon induced luminescence of GO nanoparticles. The envelope of the obtained spectrum is similar to those under one-photon and two-photon excitations. Similar to the principal of two-photon luminescence, when the GO sample is stimulated by 1215 nm laser pulses, it needs to absorb at least three-photons at the same time to get excited (Figure S5b) and then relaxed to the same lowest excited electronic-vibrational state(s) to produce luminescence emission.[16] Various three-photon emission spectra of GO nanoparticles were then obtained (Figure 2e) under fs laser excitation with different average powers, and its intensity dependence on the pump pulse energy is shown in Figure 2f. One can see that the peak emission intensity is proportional to the cubic of the pump pulse power, which verifies the characterization of a three-photon process. Although a number of papers have been reported concerning the three-photon luminescence behaviors of several materials (e.g., noble metal nanoparticles, quantum dots, and organic dyes),[18] herein we observed for the first time three-photon induced luminescence performance from GO nanoparticles. As aforementioned, the wavelength of fs laser as three-photon excitation usually locates in the range of 1000-1350 nm, which is a second optical window for *in vivo* imaging as it produces low autofluorescence. Hence, the penetration depth of multiphoton excitation light can increase, and the contrast of the signal, which is generated in deep-tissue, can be improved. Signal photobleaching can be avoided and spatial resolution of imaging can also be improved due to a higher order dependence of multiphoton absorption on laser intensity. Considering high biocompatibility and molecule loading capacity of GO nanoparticles, we believed that GO based three-photon luminescence imaging will have bright prospects in both *in vitro* and *in vivo* applications in the future.

Prior to bioimaging applications, GO nanoparticles were grafted with PEG 2000 molecules following a modified protocol previously reported. Both GO and GO-PEG samples were then characterized with FTIR spectrometer, and their spectra were shown in Figure S7. The peaks at ~2922 cm$^{-1}$ and ~1110 cm$^{-1}$ in GO-PEG sample were the signatures of C-H and C-O bonds, respectively, and their intensities are more distinct than those of GO. This illustrates that PEG 2000 molecules have been effectively conjugated with GO, as PEG molecules have very rich C-H and C-O bonds.[8c,9b] We then measured various luminescence intensities of GO nanoparticles at different pH values, and found that they kept very stable from pH 3 to 12 (Figure S8). Furthermore, previous literature has demonstrated that GO-PEG nanoparticles produced negligible cytotoxicity towards cells *in vitro*.[5b]

For cell imaging, 200 μl GO-PEG nanoparticles were added into series of cell plates with HeLa cells incubated. 2h, 6h, 24h post sample treatment, the cells were imaged with a laser scanning confocal microscope (Olympus, FV 1000) under 405 nm excitation, and the results were shown in Figure S9. As one can see, the luminescence intensity on the surface of HeLa cells increased as incubation time went by, and the image contrast also get more distinct accordingly, which illustrated that more GO-PEG nanoparticles have been uptaken by HeLa cells. Furthermore, we observed that HeLa cells were still alive, and their morphologies kept very well even after 24h-incubation of GO-PEG nanoparticles, indicating that GO can be used as a biocompatible contrast agent for bioimaging.

We then utilized the upright two-photon scanning microscope (Olympus, FV1000, fs laser at 810 nm) to perform two-photon imaging of HeLa cells, which have been incubated with GO nanoparticles for 6h. As shown in Figure S10, distinct two-photon luminescence of GO-PEG nanoparticles on cells could be observed when a fs laser power less than 10 mW was adopted. On the contrary, almost no autofluorescence was excited on the control HeLa cells (without GO-PEG incubation) with the same laser power. This indicates that the signal to noise ratio (SNR) and contrast of GO-assisted two-photon luminescence imaging can be very high. Furthermore, we also found that GO nanoparticles distributed uniformly in all cells, and it coincided very well with previously published works, which demonstrated that GO nanoparticles had a good cell-labeling capacity.[5b,8b] However, to the best of our knowledge, no one has ever investigated detailed distribution of GO nanoparticles in subcellular structures. Such an investigation may be very helpful to illustrate some important cellular actions or achieve functional manipulation. To study this, four commercially available dyes (MitoTracker (Red), ER-Tracker(Red), Golgi-RFP, Lysosomes-RFP), which can specifically target certain organelles (mitochondria, endoplasmic reticulum (ER), Golgi body, lysosome), were mixed with 200 μl GO-PEG nanoparticles and co-incubated with HeLa cells, respectively. Several hours later, the four series of treated HeLa cell samples were imaged with the upright laser scanning confocal microscope, separately. We used 810 nm-fs laser to perform two-photon luminescence imaging of GO nanoparticles, since all the four dyes have negligible one-photon absorption at 405 nm. A scanning grating in confocal microscope module was set in a wavelength range of 500-575 nm to extract emission signals from GO. For these four dyes, 543 nm-CW laser was used to perform one-photon luminescence imaging, and the scanning grating was set in a wavelength range of 600-675 nm to extract their signals, as well as to avoid signal crosstalk with one-photon luminescence of GO-PEG. The imaging results were shown in Figure 3, in which "green channel" corresponded to GO-PEG signals, "red channel" corresponded to dye signals and "overlap" was the mixture channel. Comparing the three channels, we found that almost all mitochondria, ER and lysosome could be labeled with GO-PEG nanoparticles, and most Golgi bodies were also distributed with GO-PEG nanoparticles. Recently, Xing's group investigated the subcellular distribution of PEG-CNTs, and they found PEG-CNTs were only localized in mitochondria, but not in any other subcellular components.[19] We hypothesize that the difference in the subcellular localization may mainly be attributed to the dimension of nanoparticles. According to the TEM picture in Figure 1b, the average size of our GO nanoparticles was about 40 nm, while the CNTs used in Xing's paper were hundreds of nanometer long. It might be difficult for ER, Golgi body and lysosome to be labeled with such long CNTs.

However, GO nanoparticles still remained very small even after they were conjugated with PEG molecules, and they could effectively target various subcellular components. Furthermore, functionalization of nanoparticles and properties of subcellular components could also influence the actual targeting effect of GO and CNTs. Considering the fact that GO possesses visible/NIR photoluminescence capacity while CNTs can only exhibit NIR luminescence,[20] we believe GO nanoparticles have wider cellular/subcellular applications (e.g., imaging, manipulation, molecule delivery) in the future.

In recent years, GO has been utilized in various types of *in vivo* research, such as distribution monitoring, tumor targeting, photothermal therapy, and chemotherapeutic drugs delivery.[5a,8c,21] However, no research group has ever extended their applications to animal brain imaging. In order to achieve high imaging resolution and observe the detailed information, confocal microscope rather than macro *in vivo* imaging system should be adopted. In a traditional one-photon confocal microscopy, visible excitation/emission light is prone to be absorbed by water in the tissue, and scattered due to the Rayleigh scattering effect, so the imaging depth cannot reach 100 μm. As aforementioned, by virtue of less absorption/scattering of NIR fs excitation source, deep-tissue imaging capacity is one of the advantages of two-photon luminescence microscopy. Furthermore, the nonlinear excitation mode can also help improving spatial resolution of imaging rather than that under one-photon confocal microscopy. Considering the good performance of two-photon luminescence of GO nanoparticles, here we attempted to apply them in brain imaging of live animals.

The skull of a black mouse was open up through microsurgery, and 0.2 μl GO-PEG nanoparticles were microinjected into its brain. By counting the scale on the microinjection instrument, we could know a relatively accurate XY-Z (depth) position where the GO-PEG samples located. In our experiment, the depth of injected GO-PEG nanoparticles was about 300 μm. A thin cover glass slide was mounted on the mouse brain through dental cement, and a metal ring was then adhered to the cover slide. After anesthetized with pentobarbital and immobilized with the metal ring, the mouse was imaged with the upright two-photon scanning microscope (Olympus, FV1000, fs laser at 810 nm), as shown in Figure S11. Water was smeared between a long work distance (2 mm) water immersed objective (25X, NA=1.05) and the cover slide (Figure 4a), and average power of the excitation fs laser was about 15 mW. Figure 4b shows three-dimensional reconstructed image of GO-PEG nanoparticles in the mouse brain. As one can see, the two-photon luminescence of GO nanoparticles could be discriminated from the background, and it clearly illustrated that the microinjected GO-PEG nanoparticles were distributed in a 50 μm-range around a depth of 300 μm in the mouse brain. Usually, the distribution depth of cerebellum, cerebral cortex and cerebral cortex blood vessels of small mice were not beyond 800 μm in their brains. If the photoluminescence of GO is tuned to a deep red/NIR region and the fs laser excitation condition can be further optimized, GO-based two-photon excitation can achieve such an imaging depth, and it will be very helpful to various potential *in vivo* applications (e.g., gene therapy in brain, blood brain barrier penetration).[22] Furthermore, fs laser can also assist to manipulate cellular/tissue behaviour besides multiphoton excitation. Considering the above superiority, we believe the "Nano GO-Ultrafast Laser" platform will attract huge research attentions in the future.

In summary, two-photon and three-photon induced photoluminescences of GO nanoparticles have been clearly observed, by utilizing fs pulse laser excitation. Grafted with PEG molecules, GO nanoparticles exhibited high chemical stability under various pH values, which could facilitate their *in vitro* and *in vivo* applications. Two-photon luminescence imaging clearly illustrated the distribution of GO-PEG nanoparticles in cellular/subcellular components. Since GO has been proven to have a less cytotoxicity and a better molecule-delivery capacity, while fs laser can also help manipulate cellular behaviours (e.g., cell fusion, gene transfection, and $Ca^{2+}$ influx control), we anticipate that many cellular/subcellular mechanisms and functions could be achieved with the present imaging technology assisted by GO and fs laser. Concerning *in vivo* imaging, the microinjected GO-PEG nanoparticles in a mouse brain could also be discriminated with two-photon luminescence microscopy, and the imaging depth could reach 300 μm or more. We also believe the "Nano GO-Ultrafast Laser" platform will attract huge research attentions in applications of gene therapy in brain, blood brain barrier penetration, etc. Furthermore, three-photon induced photoluminescence of GO nanoparticles has also been observed in our experiment. Since the fs laser wavelength for three-photon excitation usually locates in a second optical window for *in vivo* imaging, which is 1000-1350 nm, high imaging contrast and low autofluorescence can be achieved simultaneously. GO based three-photon luminescence technology will have bright prospects in *in vivo* imaging.

This work was partially supported by the Science and Technology Department of Zhejiang Province, the National Basic Research Program (973) of China (2011CB503700), the Special Financial Grant from the China Postdoctoral Science Foundation (No. 201104741), the National Natural Science Foundation of China (61008052 and 60990322), and the Fundamental Research Funds for the Central Universities. We started this independent work about ten months ago. Just before the submission of the present manuscript, we noticed a related work, which is on two-photon (but not three-photon) bio-imaging (but not for in vivo or subcellular components), has just been published in this journal (vol.51, 1-6, 2012).

## References

[1] K. S. Novoselov, A. K. Geim, S. V. Morozov, D. Jiang, Y. Zhang, S. V. Dubonos, I. V. Grigorieva, A. A. Firsov, *Science* **2004**, *306*, 666.

[2] a) K. S. Novoselov, A. K. Geim, S. V. Morozov, D. Jiang, M. I. Katsnelson, I. V. Grigorieva, S. V. Dubonos, A. A. Firsov, *Nature* **2005**, *438*, 197; b) C. Lee, X. D. Wei, J. W. Kysar, J. Hone, *Science* **2008**, *321*, 385.

[3] a) Z. Sun, T. Hasan, F. Torrisi, D. Popa, G. Privitera, F. Wang, F. Bonaccorso, D. M. Basko, A. C. Ferrari, *ACS Nano* **2010**, *4*, 803; b) X. Ling, L. Xie, Y. Fang, H. Xu, H. Zhang, J. Kong, M. S. Dresselhaus, J. Zhang, Z. Liu, *Nano Lett.* **2010**, *10*, 553; c) X. Li, Y. Zhu, W. Cai, M. Borysiak, B. Han, D. Chen, R. D. Piner, L. Colombo, R. S. Ruoff, *Nano Lett.* **2009**, *9*, 4359.

[4] a) K. P. Loh, Q. Bao, G. Eda, M. Chhowalla, *Nat. Chem.* **2010**, *2*, 1015; b) L. Feng, Z. Liu, *Nanomedicine* **2011**, *6*, 317.

[5] a) X. Zhang, J. Yin, C. Peng, W. Hu, Z. Zhu, W. Li, C. Fan, Q. Huang, *Carbon* **2011**, *49*, 986; b) C. Peng, W. Hu, Y. Zhou, C. Fan Q. Huang, *Small* **2010**, *6*, 1686.

[6] a) J. Qian, K. Yong, I. Roy, T. Y. Ohulchanskyy, E. J. Bergey, H. H. Lee, K. Tramposch, S. He, A. Maitra, P. N. Prasad, *J. Phys. Chem. B* **2007**, *111*, 6969; b) J. Qian, L. Jiang, F. Cai, D. Wang, S. He, *Biomaterials* **2011**, *32*, 1601; c) Q. Zhan, J. Qian, H. Liang, G. Somesfalean, D. Wang, S. He, Z. Zhang, S. Andersson-Engels, *ACS Nano* **2011**, *5*, 3744.

[7] S. Shukla, S. Saxena, *Appl. Phys. Lett.* **2011**, *98*, 073104.

[8] a) X. Sun, Z. Liu, K. Welsher, J. T. Robinson, A. Goodwin, S. Zaric, H. Dai, *Nano Res.* **2008**, *1*, 203; b) Z. Liu, J. T. Robinson, X. Sun, H. Dai, *J. Am. Chem. Soc.* **2008**, *130*, 10876; c) K. Yang, S. Zhang, G. Zhang, X. Sun, S. T. Lee, Z. Liu, *Nano Lett.* **2010**, *10*, 3318.

[9] a) G. Eda, Y. Y. Lin, C. Mattevi, H. Yamaguchi, H. A. Chen, I. S. Chen, C. W. Chen, M. Chhowalla, *Adv. Mater.* **2010**, *22*, 505–509; b) J. Chen, X. Yan, *Chem. Commun.* **2011**, *47*, 3135; c) Z. Liu, X. Zhao, X. Zhang, X. Yan, Y. Wu, Y. Chen, J. Tian, *J. Phys. Chem. Lett.* **2011**, *2*, 1972; d) Z. Liu, Y. Wang, X. Zhang, Y. Xu, Y. Chen, J. Tian, *Appl. Phys. Lett.* **2009**, *94*, 021902.

[10] a) S. Kim, H. E. Pudavar, A. Bonoiu, P. N. Prasad, *Adv. Mater.* **2007**, *19*, 3791; b) H. Wang, T. B. Huff, D. A. Zweifel, W. He, P. S. Low, A. Wei, J. X. Cheng, *P. Natl. Acad. Sci.* **2005**, *102*, 15752; c) K. T. Yong, J. Qian, I. Roy, H. H. Lee, E. J. Bergey, K. M. Tramposch, S. He, M. T. Swihart, A. Maitra, P. N. Prasad, *Nano Lett.* **2007**, *7*, 761.

[11] a) G. S. He, Q. Zheng, K. T. Yong, A. I. Ryasnyanskiy, P. N. Prasad, A. Urbas, *Appl. Phys. Lett.* **2007**, *90*, 181108; b) G. S. He, Q. Zheng, K. T. Yong, F. Erogbogbo, M. T. Swihart, P. N. Prasad, *Nano Lett.* **2008**, *8*, 2688.

[12] a) A. Yodh, B. Chance, *Phys. Today.* **1995**, *48*, 34; b) A. M. Smith, M. C. Mancini, S. Nie, *Nat. Nanotechnol.* **2009**, *4*, 710.

[13]  a) H. He, K. T. Chan, S. K. Kong, R. K. Y. Lee, *Appl. Phys. Lett.* **2009**, *95*, 233702; b) H. He, K. T. Chan, S. K. Kong, R. K. Y. Lee, *Appl. Phys. Lett.* **2008**, *93*, 163901; c) H. He, S. K. Kong, R. K. Y. Lee, Y. K. Suen, K. T. Chan, *Opt. Lett.* **2008**, *33*, 2961; d) H. He, S. K. Kong, K. T. Chan, *J. Biomed. Opt.* **2010**, *15*, 059803.

[14]  D. Li, M. B. Müller, S. Gilje, R. B. Kaner, G. G. Wallace, *Nat. Nanotechnol.* **2008**, *3*, 101.

[15]  K. N. Kudin, B. Ozbas, H. C. Schniepp, R. K. Prud'homme, I. A. Aksay, R. Car, *Nano Lett.* **2008**, *8*, 36.

[16]  G. S. He, L. S. Tan, Q. Zheng, P. N. Prasad, *Chem. Rev.* **2008**, *108*, 1245.

[17]  C. Xu, W. Zipfel, J. B. Shear, R. M. Williams, W. W. Webb, *Proc. Natl. Acad. Sci.* **1996**, *93*, 10763.

[18]  a) L. Tong, C. M. Cobley, J. Chen, Y. Xia, J. X. Cheng, *Angew. Chem. Int. Ed.* **2010**, *49*, 3485; b) C. Lu, W. Huang, J. Luan, Z. Lu, Y. Qian, B. Yun, G. Hu, Z. Wang, Y. Cui, *Optics Commun.* **2008**, *281*, 4038; c) G. S. He, K. T. Yong, Q. Zheng, Y. Sahoo, A. Baev, A. I. Ryasnyanskiy, P. N. Prasad, *Opt. Expr.* **2007**, *15*, 12818.

[19]  F. Zhou, D. Xing, B. Wu, S. Wu, Z. Ou, W. R. Chen, *Nano Lett.* **2010**, *10*, 1677.

[20]  a) Z. Luo, P. M. Vora, E. J. Mele, A. T. C. Johnson, J. M. Kikkawa, *Appl. Phys. Lett.* **2009**, *94*, 111909; b) Z. Liu, S. Tabakman, K. Welsher, H. Dai, *Nano Res.* **2009**, *2*, 85.

[21]  L. Zhang, J. Xia, Q. Zhao, L. Liu, Z. Zhang, *Small* **2010**, *6*, 537.

[22]  a) A. C. Bonoiu, S. D. Mahajan, H. Ding, I. Roy, K. T. Yong, R. Kumar, R. Hu, E. J. Bergey, S. A. Schwartz, P. N. Prasad, *Proc. Natl. Acad. Sci.* **2009**, *106*, 5546; b) D. J. Bharali, I. Klejbor, E. K. Stachowiak, P. Dutta, I. Roy, N. Kaur, E. J. Bergey, P. N. Prasad, M. K. Stachowiak, *Proc. Natl. Acad. Sci.* **2005**, *102*, 11539.

**Figures and Figure Captions:**

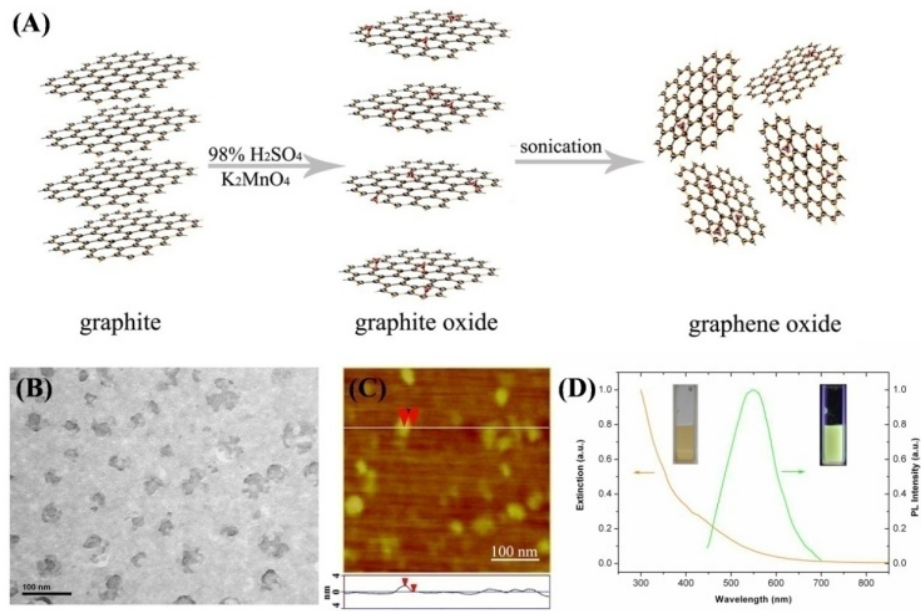

*Figure 1.* Synthesis and characterization of our GO nanoparticles. a) scheme illustrating the synthesis process of GO nanoparticles from bulk graphite; b) a TEM picture of GO nanoparticles; c) an AFM image of GO nanoparticles; d) extinction and one-photon luminescence spectra of aqueous solution of GO nanoparticles (Inset: pictures of an aqueous solution of GO nanoparticles under daylight lamp and UV lamp).

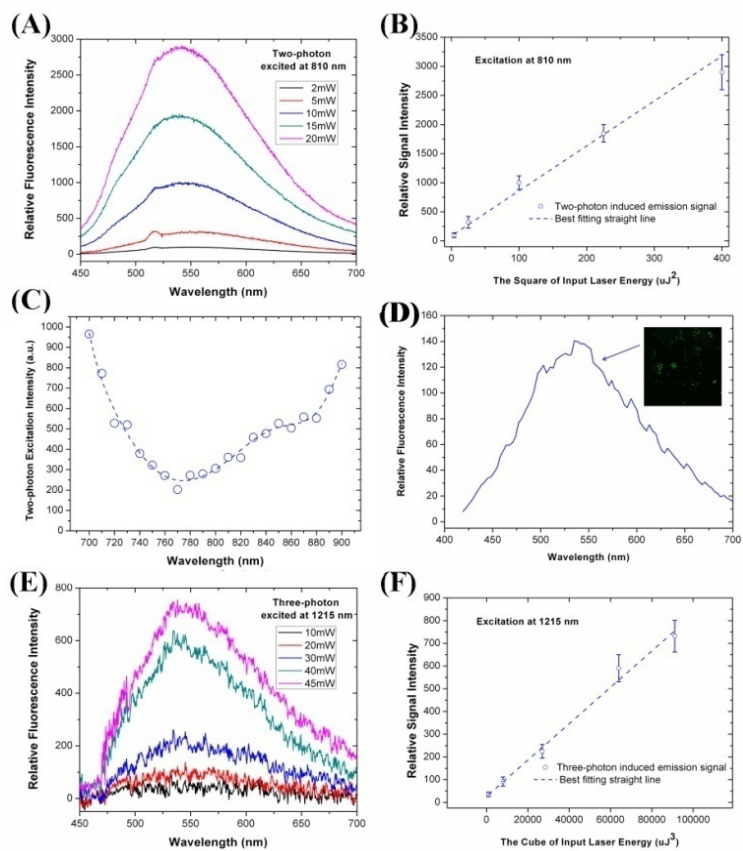

*Figure 2.* Characterization of multiphoton luminescence of GO nanoparticles. a) two-photon luminescence spectra of GO nanoparticles under various average powers of fs laser excitation; b) square dependence of two-photon luminescence intensity on excitation pulse energy of fs laser; c) fitted average powers required for fs laser of different wavelengths to achieve the same two-photon luminescence intensity of GO nanoparticles; d) two-photon luminescence spectrum and image (inset) of a diluted aqueous solution of GO nanoparticles; e) three-photon luminescence spectra of GO nanoparticles under various average powers of fs laser excitation; f) cubic dependence of three-photon luminescence intensity on excitation pulse energy of fs laser.

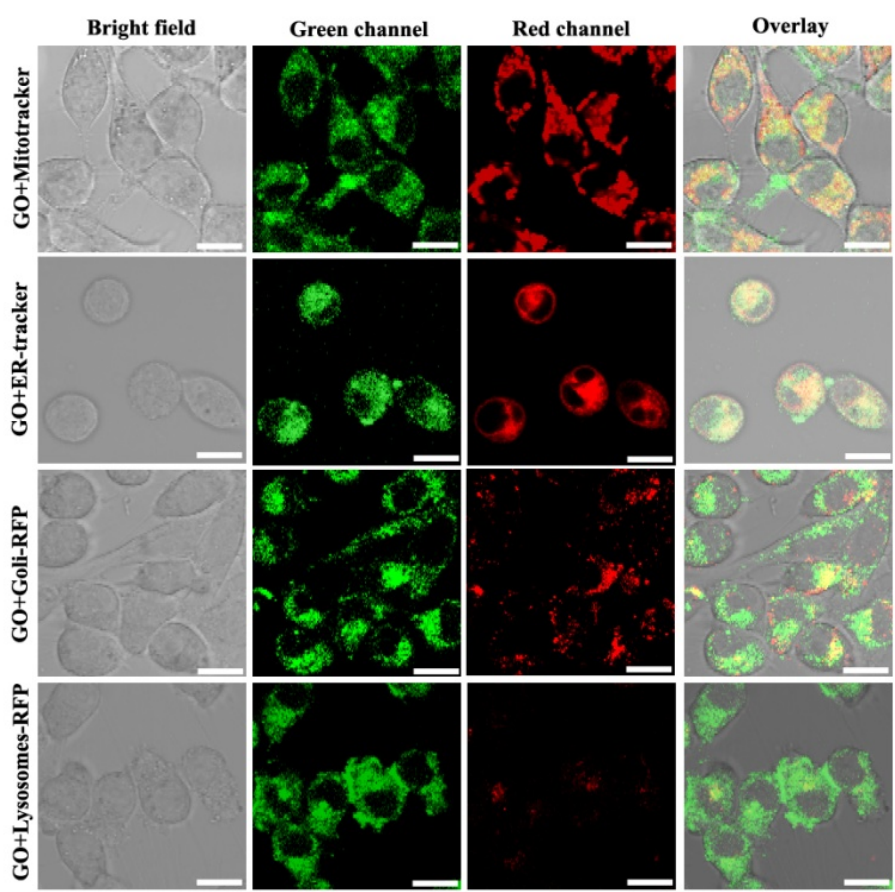

*Figure 3.* Localization of GO-PEG nanoparticles in various subcellular components (mitochondria, ER, Golgi body, lysosome) of HeLa cells, which was illustrated by using two-photon luminescence microscopy. Scale bar: 20 μm.

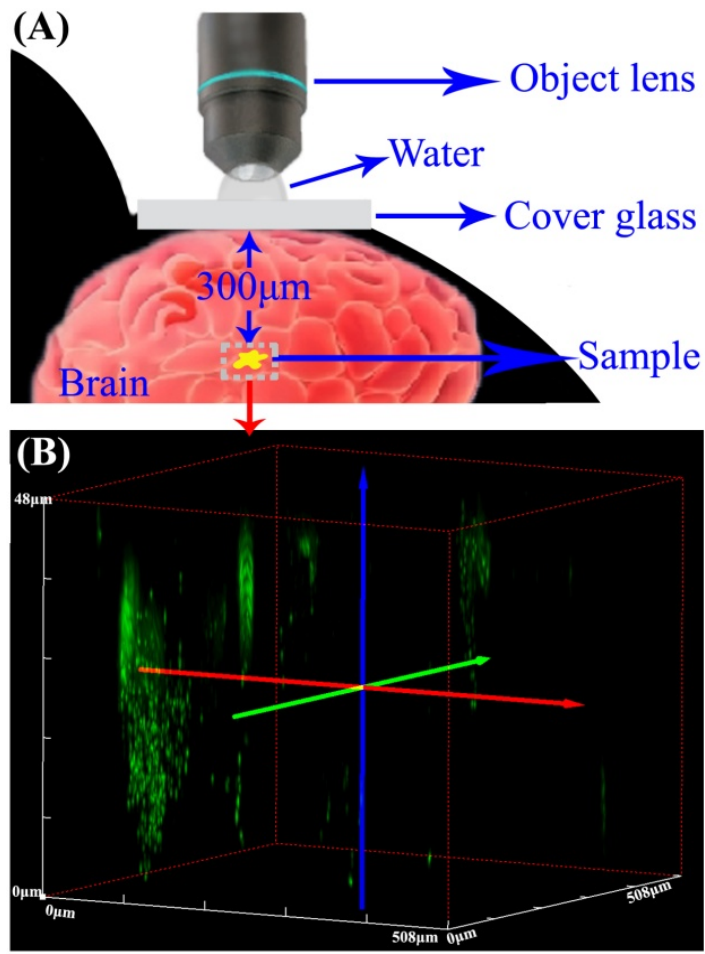

***Figure 4.*** Two-photon luminescence imaging of GO nanoparticles in a mouse brain. a) scheme illustrating the two-photon luminescence microscopy of a mouse brain; b) reconstructed image of the three-dimensional distribution of GO-PEG nanoparticles.

# Supporting information

*Experimental Section*

**Materials.** Graphite, sodium chloride (NaCl), concentrated sulfuric acid ($H_2SO_4$, 98%), potassium permanganate ($KMnO_4$), hydrogen peroxide ($H_2O_2$, 30%), hydrochloric acid (HCl, 36%), PEG 2000, 1,1-carbonyldiimidazole (CDI) and dimethyl sulphoxide (DMSO) were obtained from Sigma-Aldrich Co. Cell-culture products, as well as MitoTracker (Red), ER-Tracker(Red), Golgi-RFP, Lysosomes-RFP, were purchased from Invitrogen Co. All the reagents were used without further purification, and deionized (DI) water was used in all the experimental procedures.

**Synthesis of graphene oxide (GO) nanoparticles.** GO nanoparticles were synthesized from natural graphite with a modified Hummer's method.[1] In brief, 1 g graphite was ground with 40 g NaCl for 10 minutes by using mortar and pestle. NaCl was dissolved in DI water and removed by filtration, and the ground graphite flakes were then added to 30 ml $H_2SO_4$ (98%) and left stirring for 10 h. Afterwards, 6 g $KMnO_4$ was added to the solution while keeping the temperature under 10ºC. The solution was stirred at 40 ºC for 30 minutes, and then stirred at 90 ºC for 90 minutes. Next, 50 ml DI water was added, and the heat temperature was increased to 100ºC. After 30 minutes, additional 100 ml $H_2O$ was added to dilute the solution, and 30 mL 30% $H_2O_2$ was injected into the solution to completely react with the excess $KMnO_4$. For purification, the resulting mixture was washed multiple times, first with 5% HCl solution and then with DI water. The obtained graphite oxide was dispersed in water and sonicated for 6 hours. After that, the solution was filtrated by amicroporous membrane (0.22 μm). The filtrate containing GO nanoparticles was collected and stored for further use.

**Preparation of GO-PEG 2000.** GO-PEG conjugates were prepared following a modified protocol previously reported by Peng et al.[2] Firstly, 5 ml aqueous solution of GO nanoparticles at a concentration of ~1 mg/ml and 5 mg CDI were allowed to react in DMSO (50 mL) at 40 ºC for 2 h with strong stirring. The resulting suspension was spin-filtered using a microfuge membrane-filter (NANOSEP 100KOMEGA,

Pall Corporation, USA) at 13,000 rpm for 30 min to remove the excess CDI. The precipitate was resuspended in DMSO (10 ml) and sonicated for about 30 min to form a clearsolution. Secondly, 5 mg of PEG 2000 was added to the CDI-activated GO suspension and the mixed solution was stirred for 24 h at 40 ºC. Finally, the reacted solution was dialyzed in a 5 kDa cutoff cellulose membrane for several days to remove DMSO and excess PEG 2000. The obtained aqueous solution of GO-PEG nanoparticles was stored at 4 ºC for further study.

**Characterization of GO nanoparticles.** AFM images were recorded using a MultiMode scanning probe microscopes (Veeco, USA) in a tapping mode with a scanning rate of 1 Hz. A droplet of GO dispersion was cast onto a freshly cleaved mica surface, and was then measured after dried at room temperature. TEM images were taken by a JEOL JEM-1230 transmission electron microscope operating at 160 kV in bright-field mode. Raman spectra of GO nanoparticles solution were measured by a BWTEK Raman probe. A 785nm excitation laser light with a power of 300 mW was introduced by a fiber to the sample, and the collected Raman signal (integration time: 5 s) was sent through another fiber to the Raman spectrometer. FTIR spectra of GO and GO-PEG nanoparticles were measured by a Thermo Nicolet FTIR spectrometer. Extinction and linear transmission spectra of GO nanoparticles were recorded by a Shimadzu 2550 UV-vis scanning spectrophotometer. One-photon luminescence and excitation spectra were obtained by a Fluorescence Spectrophotometer (F-2500, HITACHI, Japan).

**Two-photon luminescence microscopy.** An upright Olympus laser scanning confocal microscope (FV1000) was used for both one-photon and two-photon luminescence imaging. When standard one-photon imaging was performed, CW lasers with various wavelengths (405, 488, 515, 543, 633 nm) were utilized as excitation, and a confocal pinhole was adopted to achieve high spatial resolution. For two-photon imaging, a fs Ti:sapphire laser (MaiTai HP, Spectra Physics) with a repetition rate of 80 MHz and a tunable wavelength from 690 to 1040 nm was used as excitation, and the confocal pinhole was not adopted since confinement of nonlinear excitation could also achieve high spatial resolution (The former technology was called laser scanning confocal microscopy, and the latter was called scanning two-photon luminescence microscopy). To image GO nanoparticles, original or diluted solution of GO sample was added on a glass slide, and a cover slide with spacers was then covered directly on the sample. A long work distance (3.3 mm) water immersed objective (40X, NA=0.8) was adopted for imaging, and water was smeared on the cover slide to match their refractive indices. Two-photon imaging of GO nanoparticles was performed under 810 nm-fs laser excitation, and corresponding luminescence spectra were recorded with an optical fiber spectrometer (PG2000, Ideaoptics Instruments).

**In vitro cell imaging.** HeLa cells (human cervical carcinoma cell lines) were cultivated in Dulbecco minimum essential media (DMEM) with 10% fetal bovine serum (FBS), 1% penicillin, and 1% amphotericin B. One day before the treatment, the cells were seeded in 35 mm cultivation dishes at a confluence of 70-80%. During the treatment, 200 μl stock solution of GO-PEG nanoparticles was added into each HeLa cell plate. HeLa cell plates without any treatment were used for control experiment. The cell incubation process lasted for 2, 6, 24 hours at 37°C with 5% $CO_2$. Then the cells were washed thrice with PBS (phosphate buffered saline, 1X) and imaged with the upright Olympus laser scanning confocal microscope. 405 nm-CW laser was used for one-photon imaging, while 810 nm-fs laser was used for two-photon imaging. The 40X objective (water immersed) was inserted into PBS solution in cell plates during imaging.

**Concerning In vivo animal imaging.** All *in vivo* experiments were performed in compliance with Zhejiang University Animal Study Committee's requirements for the care and use of laboratory animals in research. 18-21 g black mice (C57 line) were used for brain imaging studies. The animal housing area (located in Animal Experimentation Center of Zhejiang University) was maintained at 24 °C with a 12 h light/dark cycle, and animals were fed with water and standard laboratory chow.

**Reference**

[1] W. S. Hummers, R. E. Offeman, *J. Am. Chem. Soc.* **1958**, *80*, 1339-1339.

[2] C. Peng, W. Hu, Y. Zhou, C. Fan, Q. Huang, *Small* **2010**, *6*, 1686.

# SUPPORTING FIGURES

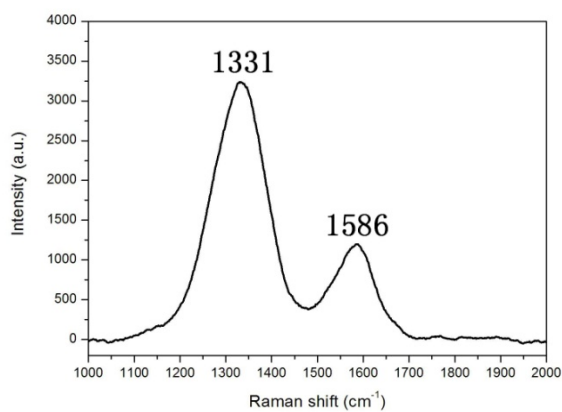

**Figure S1.** Enhanced Raman spectrum of GO nanoparticles.

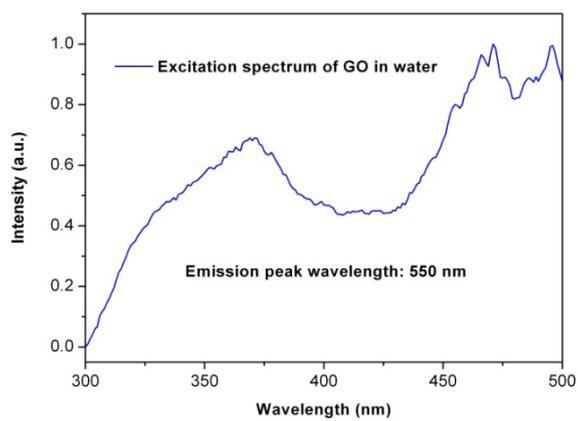

**Figure S2.** Average power intensity required for excitation light of different wavelengths to achieve one-photon luminescence with the same intensity at the peak wavelength of 550 nm for an aqueous solution of GO nanoparticles.

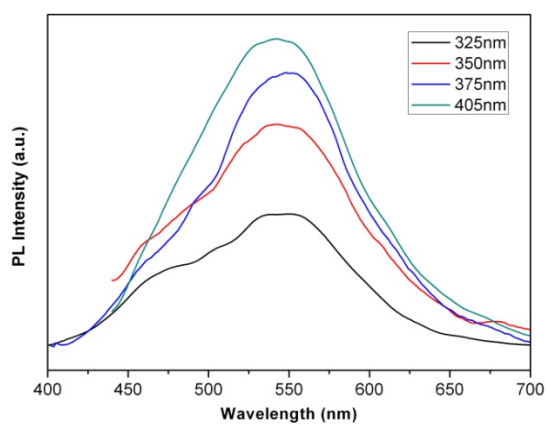

**Figure S3.** One-photon luminescence spectra of GO nanoparticles under various excitation wavelengths.

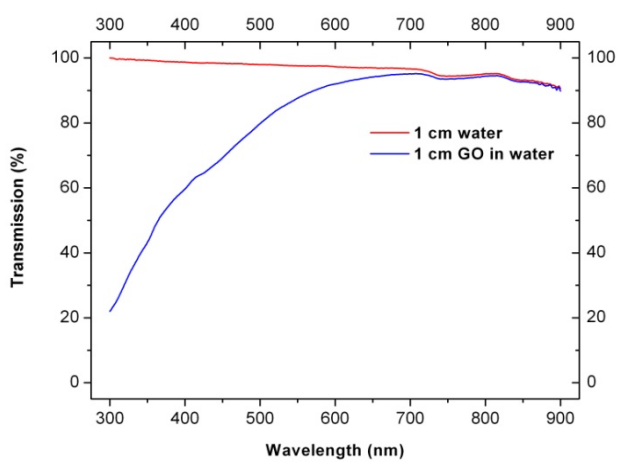

**Figure S4.** Linear transmission spectra of 1-cm-thick layer of water and aqueous dispersion of GO nanoparticles.

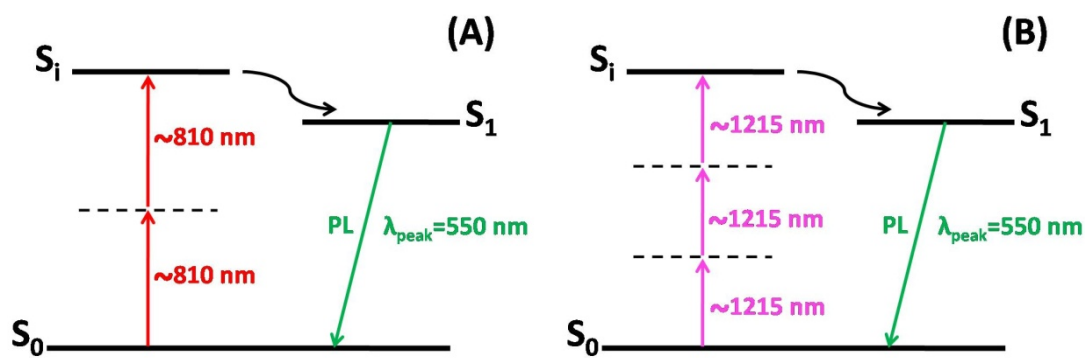

**Figure S5.** Diagrams showing the proposed mechanism for two-photon and three-photon excitation and luminescence. $S_i$ is a higher (electronic or vibronic) state; $S_1$ is the lower radiative state; $S_0$ is the ground state. Under our experimental conditions, luminescence occurred as a result of radiative transition between $S_1$ and $S_0$.

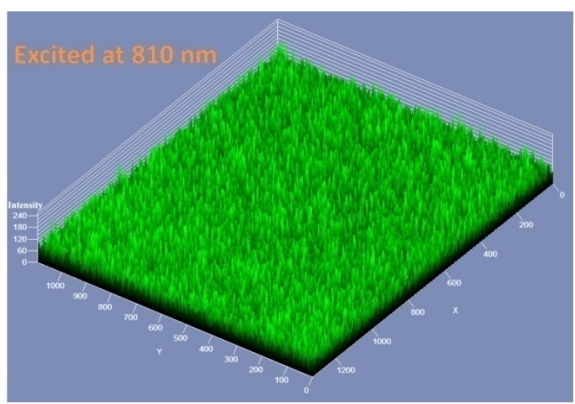

**Figure S6.** A 3D reconstructed picture illustrating the intensity dependence on the X-Y position of GO nanoparticles. Excitation wavelength of fs laser: 810 nm.

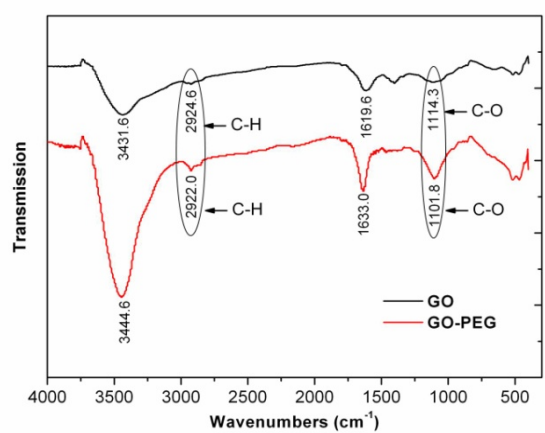

**Figure S7.** FTIR spectra of GO and GO-PEG nanoparticles.

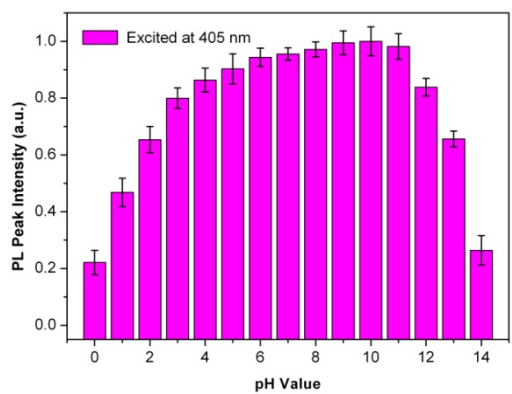

**Figure S8.** Normalized luminescence intensity of GO-PEG under various pH values.

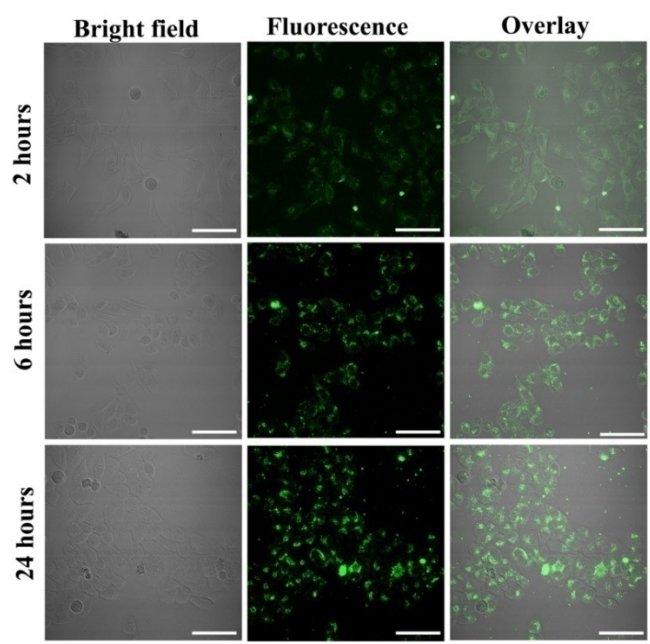

**Figure S9.** One-photon luminescence confocal images of HeLa cells, 2h, 6h and 24h post treatment of GO-PEG nanoparticles. Scale bar: 50 μm.

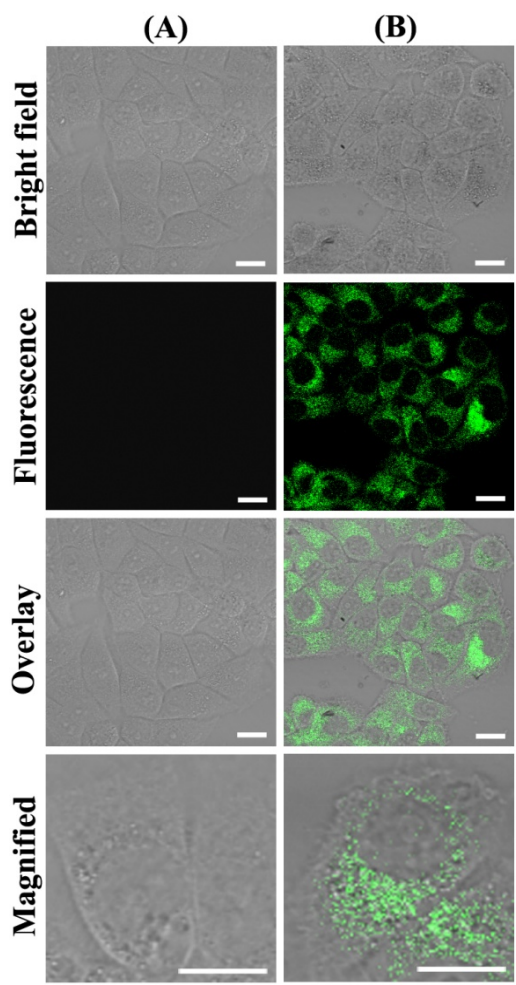

**Figure S10.** Two-photon luminescence confocal images of HeLa cells, 6h post treatment of GO-PEG nanoparticles. Scale bar: 20 μm.

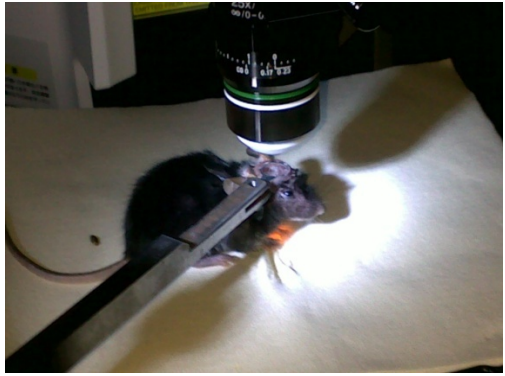

**Figure S11.** A picture showing the imaging processof a mouse brain under the two-photon scanning microscope.